\documentclass[12pt]{iopart}
\usepackage{iopams}
\usepackage{graphicx} % Required for inserting images
\usepackage{amsmath}	% Include figure files
\usepackage{hyperref}
\hypersetup{colorlinks=true, urlcolor=cyan, linkcolor=blue, citecolor=blue}
\usepackage{xspace}	% Include xspace

\newcommand{\bjet}{\mbox{$b$-jet}\xspace}
\newcommand{\bjets}{\mbox{$b$-jets}\xspace}
\newcommand{\pythia}{\mbox{\textsc{Pythia8}}\xspace}
\newcommand{\delphes}{\mbox{\textsc{Delphes}}\xspace}
\newcommand{\ptjet}{\mbox{$p_{T,~\mathrm{jet}}$}\xspace}
\newcommand{\etajet}{\mbox{$\eta_\mathrm{jet}$}\xspace}
\newcommand{\phijet}{\mbox{$\phi_\mathrm{jet}$}\xspace}
\newcommand{\mjet}{\mbox{$m_\mathrm{jet}$}\xspace}
\newcommand{\db}{\mbox{$D_b$}\xspace}
\newcommand{\PbPb}{\mbox{Pb--Pb}\xspace}
\newcommand{\raa}{\mbox{$R_{AA}$}\xspace}

\newcommand{\ptrm}{\mbox{$p_{\rm T}$}\xspace}

\begin{document}

\title{Investigation of performance of a GNN-based $b$-jet tagging method in heavy-ion collisions}

\newcommand{\pusan}{Department of Physics, Pusan National University, Busan 46241, South Korea}
\newcommand{\pusanepi}{Extreme Physics Institute, Pusan National University, Busan 46241, South Korea}

\author{Changhwan Choi} \address{\pusan} \ead{changhwan.choi@cern.ch}
\author{Sanghoon Lim} \address{\pusan}\address{\pusanepi} \ead{shlim@pusan.ac.kr}

%\author{Content \& Services Team}
%\address{IOP Publishing, No. 2 The Distillery, Avon Street, Bristol BS2 0GR, UK}
%\ead{customerservices@ioppublishing.org}
\vspace{10pt}
%\begin{indented}
%\item[]August 2017 (minor update March 2024)
%\end{indented}

\begin{abstract}

Beauty-tagged jets ($b$-jets)—collimated sprays of particles originating from the fragmentation of beauty quarks produced in the initial hard scatterings—provide a unique probe of parton dynamics in the quark-gluon plasma (QGP) created in ultrarelativistic heavy-ion collisions. In particular, energy loss patterns of low-$p_T$ $b$-jets traversing the QGP offer valuable insight into the strong interaction in its non-perturbative regime. CMS and ATLAS Collaborations at the LHC have studied $b$-jet production in Pb--Pb collisions. The results were limited to a high-$p_{\rm T}$ region, because a major challenge at low-$p_{\rm T}$ is the overwhelming number of background particles from QGP hadronisation, which severely hinders the effectiveness of conventional $b$-jet tagging techniques. To enable precise measurements in such complex environments, advanced tagging methods are required. Graph Neural Networks (GNNs), capable of learning relational structures among jet constituents, represent a promising deep learning approach for $b$-jet identification. In this study, we adopt and adapt the GN1 model, initially developed by ATLAS, for use in Pb–Pb collision environments. We investigate the model’s performance by applying it to jets embedded with Pb–Pb background particles, evaluating both tagging decisions and robustness against background contamination. This work presents a comprehensive evaluation of GNN-based $b$-jet tagging under heavy-ion collision conditions, aiming to advance future precision studies of QGP-induced partonic energy loss.

\end{abstract}

%\maketitle

\pagebreak

\section{Introduction}
\label{sec:Intro}

Beauty-tagged jets (\bjets) are high-momentum, collimated sprays of particles that contain a beauty quark or hadron. These jets typically originate from the fragmentation of beauty quarks ($b$), produced via initial hard scatterings or gluon splitting. While jets can also emerge from hard-scattered light quarks ($u$, $d$, $s$), gluons ($g$), or charm quarks ($c$), heavy-flavour jets like \bjets are of particular interest in Quark–Gluon Plasma (QGP)~\cite{Busza:2018rrf} studies due to the large mass of the beauty quark ($m_b = 4.183 \pm 0.007~\text{GeV}/c^2$~\cite{PDG}). This mass is significantly greater than both the QCD scale and the typical QGP temperature at the LHC, implying that $b$ quarks are predominantly produced in the early stages of collisions through perturbative QCD processes. Their well-defined production mechanisms make \bjets ideal probes for studying the transport properties of the QGP via both radiative and collisional energy loss mechanisms (for a recent review, see Ref.~\cite{Dong:2019byy}).

As jets traverse the QGP medium formed in heavy-ion collisions, they lose energy in a process known as jet quenching. The difference in quenching between \bjets and inclusive jets reveals essential QCD phenomena, particularly due to the mass dependence of parton energy loss. For heavy-flavour quarks, soft gluon radiation is suppressed at small angles ($\theta < m/E$) relative to their direction of motion—a phenomenon known as the dead cone effect~\cite{Dokshitzer:2001zm, ALICEdeadcone}. Consequently, heavy-flavour quarks lose less energy through medium-induced radiation compared to light partons and experience less quenching~\cite{Zigic:2018ovr, Wicks:2005gt, Ke:2018tsh}.

Previous measurements of the nuclear modification factor \raa for \bjets in \PbPb collisions by CMS and ATLAS at $\sqrt{s_\mathrm{NN}}=2.76$ and 5.02 TeV, respectively~\cite{ATLASbjetRAA, CMSbjetRAA}, were limited to high transverse momentum ($p_\mathrm{T,~jet} > 80~\mathrm{GeV}/c$). In this regime, the observed differences between \bjets and inclusive jets are predominantly attributed to colour charge effects due to differing quark and gluon jet fractions. These measurements were not able to clearly isolate mass-dependent energy loss or collisional dissipation effects. To probe such mechanisms, it is crucial to explore the low-\ptrm regime, where these effects are expected to be more pronounced. In this context, the ALICE detector offers a significant advantage, with its excellent tracking efficiency and resolution for low-\ptrm particles~\cite{ALICEreview, ALICEupgrade}.

Nevertheless, identifying \bjets in heavy-ion collisions remains challenging. The high particle multiplicity in such events—arising from QGP hadronisation and multi-parton interactions—leads to substantial background, which can obscure jet substructure and degrade the performance of classical tagging methods~\cite{CMS:2012feb, CMSJPPbPb, ATLAS:2015thz, ALICEbjet}. This motivates the need for a novel tagging approach, optimised for the ALICE environment and capable of operating effectively in high-background conditions, particularly to enable precision measurements of low-\ptrm \bjets.

In this study, we explore the use of Graph Neural Networks (GNNs) for \bjet tagging, adapting the GN1 model originally developed by the ATLAS Collaboration for pp collisions~\cite{ATLASGN1}. GNNs are deep learning architectures designed to operate on graph-structured data, where nodes represent entities (e.g., jet constituent tracks) and edges encode relationships among them~\cite{GATv2}. Since jets are composed of multiple charged tracks whose correlations carry essential flavour information—such as displaced vertices due to $b$-hadron decays—GNNs are particularly well suited to the \bjet tagging task. The GN1 model has demonstrated superior performance over both classical methods and previous deep learning models (e.g., DL1r) in ATLAS studies~\cite{Mondal:2024nsa, ATLAS:2022qxm}.

To evaluate the model’s robustness in the heavy-ion environment, we follow a two-stage procedure. First, we overlay jets from pp collisions with background particles generated from \PbPb events and apply the pretrained GN1 model (trained on pp jets) to these overlaid samples, assessing its performance degradation. Second, we retrain the GN1 model on these background-overlaid jets to investigate whether such training can improve tagging robustness under realistic \PbPb conditions. Through these investigations, we aim to assess the feasibility and performance of GNN-based \bjet tagging methods in the ALICE experiment for future heavy-ion analyses.

\section{Simulation framework}  
\label{sec:Simulation}

Simulations were carried out independently for jets from pp collisions and background particles from \PbPb collisions, both at a center-of-mass energy of $\sqrt{s_{\mathrm{NN}}} = 5.02$~TeV. The \PbPb background particles were subsequently overlaid onto the pp jets to investigate their impact on jet properties and tagging performance. Although this approach does not capture phenomena such as jet energy loss or jet shape modifications that occur in actual heavy-ion collisions, it facilitates a controlled comparison of the GNN model’s response with and without background, given identical input jets.

For the jet sample, 20 million pp collision events were generated using the \pythia Monte Carlo event generator\cite{PYTHIA8} at $\sqrt{s}=5.02$ TeV. Half of these events were used for training and validation, while the remaining half were reserved for model evaluation. To enhance the yield of events containing high-\ptrm jets, a bias was applied to the $\hat{p}_\mathrm{T}$ of hard QCD processes using a power-law weighting with an exponent of 4, over the range of $5~\mathrm{GeV}/c$ to $300~\mathrm{GeV}/c$ (\texttt{PhaseSpace:bias2SelectionPow = 4}).

The detector response was simulated using the \delphes fast simulation framework~\cite{Delphes}, which employs a parametrized approach rather than the detailed modeling provided by full simulation frameworks like \textsc{Geant4}~\cite{GEANT4:2002zbu}. While this results in reduced accuracy in reproducing complex detector effects, it enables efficient large-scale simulations with significantly lower computational overhead. In addition to track propagation and smearing, \delphes supports jet reconstruction through the \textsc{FastJet} package~\cite{FastJet}.
The simulation chain in \delphes consists of several modular stages. It begins by propagating generated particles from \pythia through a 0.5T longitudinal magnetic field, applying tracking efficiency and resolution consistent with ALICE midrapidity tracking detectors~\cite{ALICEreview, ALICEupgrade}. The Distance of Closest Approach (DCA) to the primary vertex is calculated for each track to emulate impact parameter measurements based on the performance of the ALICE ITS2.

Charged jets are reconstructed from tracks with $p_\mathrm{T} > 0.15~\mathrm{GeV}/c$ and $|\eta| < 0.9$, using the anti-$k_\mathrm{T}$ algorithm with a resolution parameter $R = 0.4$~\cite{AntikT}. Jets with $p_\mathrm{T,jet} > 10~\mathrm{GeV}/c$ and $|\eta_\mathrm{jet}| < 0.5$ are selected for further analysis. In parallel, particle-level jets are reconstructed using stable generated particles with the same algorithm and selection criteria. These are categorized into three flavour classes—\bjets, $c$-jets, and light-flavour jets ($u$, $d$, $s$, and $g$)—based on the presence of heavy quarks within $\Delta R < 0.4$ of the jet axis, where $\Delta R$ is defined as $\sqrt{(\eta_{\rm track} - \eta_{\rm jet})^{2} + (\phi_{\rm track}-\phi_{\rm jet})^{2}}$. Reconstructed jets are matched geometrically to particle-level jets, and the jet flavour is assigned accordingly.

To simulate the heavy-ion background, events were generated using the AMPT (A Multi-Phase Transport) model~\cite{AMPT}, a Monte Carlo framework that incorporates initial conditions, partonic scatterings, and hadronization processes. A total of 400,000 \PbPb events at $\sqrt{s_{\mathrm{NN}}} = 5.02$~TeV were produced. For simplicity, all background particles were assumed to originate from the primary vertex with zero impact parameter. Since AMPT does not natively output in a \delphes-compatible format, detector effects consistent with \delphes were applied during preprocessing.
The collision centrality of the AMPT events was determined using particle multiplicities in pseudorapidity regions corresponding to the ALICE V0A ($2.8 < \eta < 5.1$) and V0C ($-3.7 < \eta < -1.7$) detectors, following procedures from ALICE Run 2 analyses~\cite{ALICE:2013hur}. Events were grouped into five centrality classes: 0--20\%, 20--40\%, 40--60\%, 60--80\%, and 80--100\%.

\begin{figure}[htb]
\centering
  \includegraphics[width=0.55\linewidth]{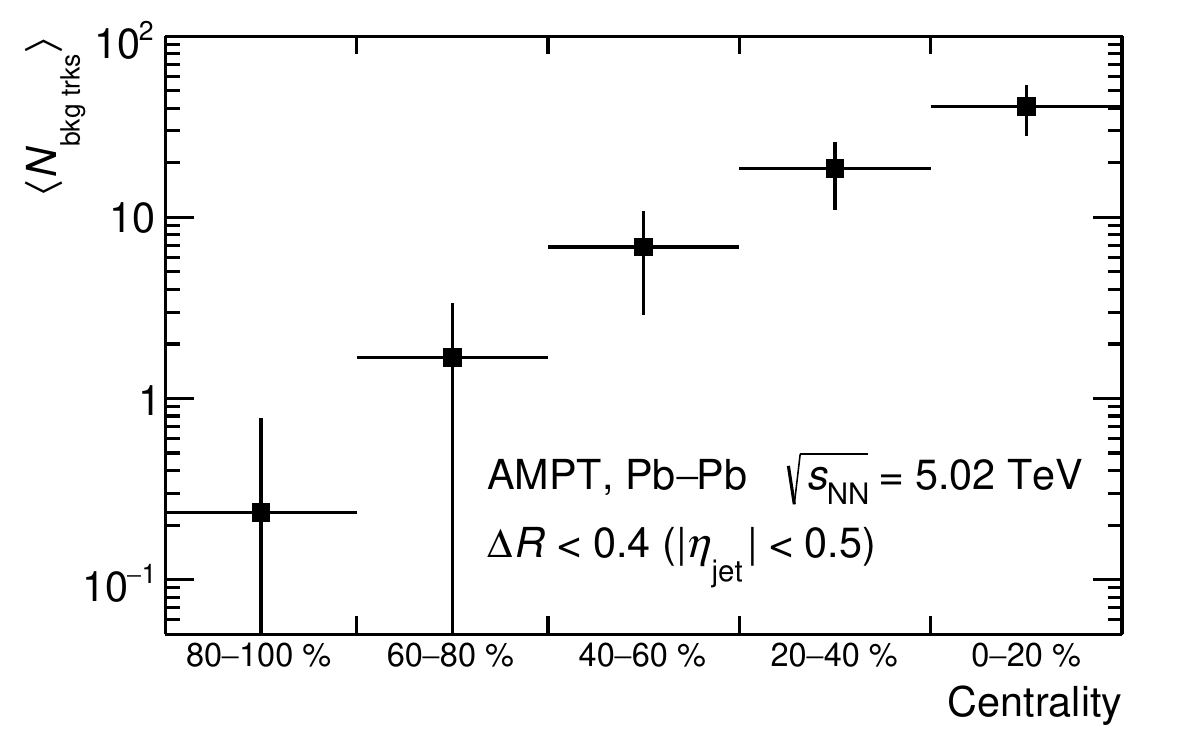}
\caption{Average number of AMPT background particles within $\Delta R < 0.4$ of jets with $|\eta_\mathrm{jet}| < 0.5$, shown as a function of collision centrality.}
\label{fig:ampt_nbkg}
\end{figure}

The procedure for overlaying a random AMPT \PbPb background event onto a given \pythia + \delphes pp jet is as follows. Background particles located within a cone of $\Delta R < 0.4$ from the jet axis are added to the list of jet constituent tracks. These background tracks are assumed to originate from the primary vertex. The momentum of the resulting background-overlaid jet is recalculated by vectorially adding the momenta of the background particles to that of the original jet, such that 
\begin{align}
    \mathbf{p}_\mathrm{jet}' = \mathbf{p}_\mathrm{jet} + \sum{\mathbf{p}_\mathrm{bkg}}.
\end{align}
Each background-overlaid jet is uniquely matched to its corresponding original pp jet to preserve one-to-one correspondence. For the input to the GNN, the kinematic properties of the jet after background overlay—namely $p_\mathrm{T,~jet}'$, $\eta_\mathrm{jet}'$, $\phi_\mathrm{jet}'$, and $m_\mathrm{jet}'$—are used. However, all performance results are reported as a function of the original transverse momentum of the pp jet, \ptjet, to enable a consistent interpretation across background conditions.

\section{Graph neural networks}
\label{sec:GNNs}

\subsection{Architecture}
\label{sec:GNNs_Arch}

\begin{figure*}[htb]
    \centering
    \includegraphics[width=0.9\linewidth]{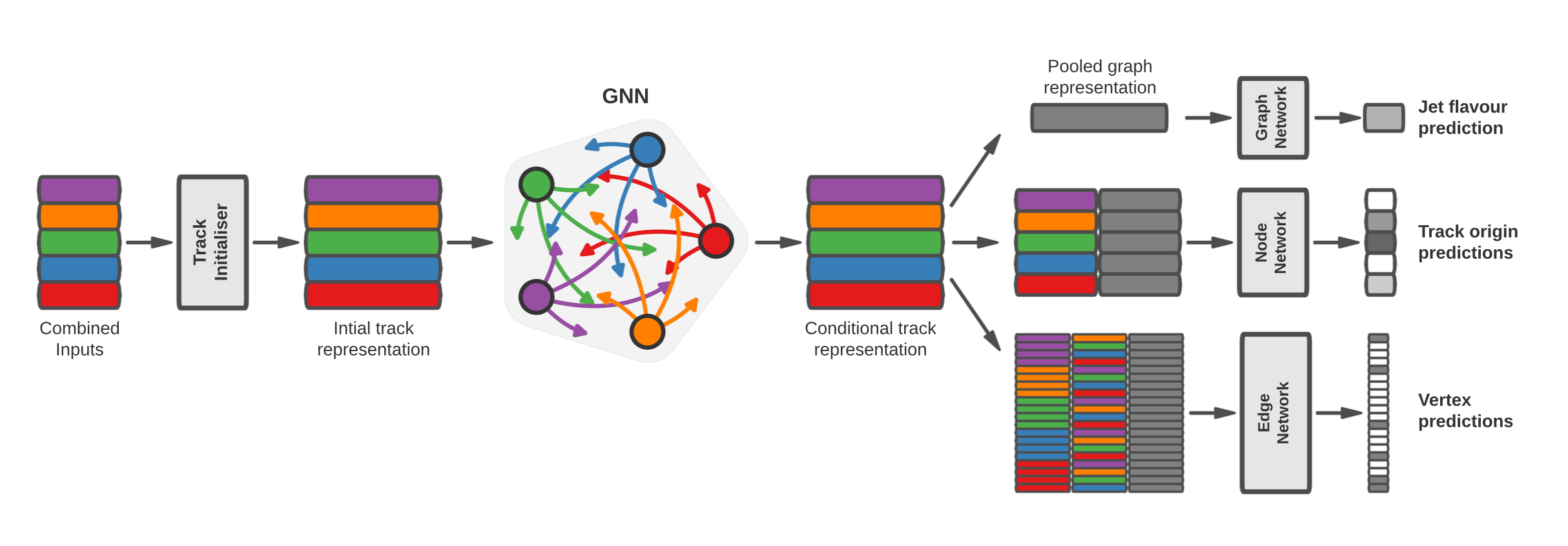}
    \caption{The architecture of ATLAS GN1~\cite{ATLASGN1}.}
    \label{fig:ATLASGN1}
\end{figure*}

The model used in this study is a modified version of the GN1 $b$-jet tagging algorithm~\cite{ATLASGN1}, a GNN-based architecture originally developed by the ATLAS Collaboration using the PyTorch and PyTorch Lightning libraries~\cite{PyTorch, PyTorchL}. The input features, prediction labels, and data input/output formats have been revised or newly implemented to suit the variables for the ALICE experiment. However, the core network architecture, including the number of layers, loss function, optimizer, and other hyperparameters, follows the specifications of the original GN1 model.

The input to the network includes both jet-level and track-level physical properties. These features consist of the momentum of the jet and its constituent tracks, as well as the track impact parameters. The overall feature set closely follows that used in the original ATLAS GN1 model, with some variables adapted to match the ALICE detector configuration. A complete list of input features is provided in Table~\ref{tab:gnn_input_pbpb}.
For GNN input, only constituent tracks with $p_\mathrm{T} > 0.5~\mathrm{GeV}/c$ are considered. Jets without any such tracks are excluded from the $b$-jet selection. If a jet contains more than 40 constituent tracks, only the top 40 tracks ranked by impact parameter significance are retained. The input data is represented as a fully connected graph, where each node corresponds to a track and includes concatenated jet-level features, enabling the network to learn attention across all possible pairwise relationships among tracks.

\begin{table}[!htb]
\centering
\begin{tabular}{ll}
\hline
\textbf{Jet input} & \textbf{Description} \\
\hline
\ptjet & Charged-particle jet transverse momentum \\
\etajet & Jet pseudorapidity \\
\phijet & Jet azimuthal angle \\
\mjet & Jet invariant mass \\
\hline
\textbf{Track input} & \textbf{Description} \\
\hline
\ptrm & Track transverse momentum \\
$\eta$ & Track pseudorapidity \\
$\phi$ & Track azimuthal angle \\
$q$ & Track charge \\
$d_{xy}$ & Signed transverse impact parameter \\
$d_{z}$ & Signed longitudinal impact parameter \\
\hline
\end{tabular}
\caption{Input features to the GNN model for the simulation study.}
\label{tab:gnn_input_pbpb}
\end{table}

% Output (predictions)
The model architecture comprises an initialisation module, a core GNN that encodes the joint features of the jet and its constituents, and three output heads designed for separate prediction tasks: (1) primary jet flavour classification, and (2) two auxiliary tasks aimed at improving tagging performance.
The primary task is jet flavour classification, which determines whether a given jet is a $b$-jet, $c$-jet, or light-flavour jet. The network outputs class probabilities $p_b$, $p_c$, and $p_\mathrm{lf}$ for each flavour via a softmax activation, such that $\sum_{i \in {b,c,\mathrm{lf}}} p_i = 1$ and $0 < p_i < 1$. These outputs are used to perform the $b$-jet tagging.

The first auxiliary task is track origin classification, which predicts the physical origin of each constituent track among five categories. The model produces per-track probability distributions over the origin classes, as defined in Table~\ref{tab:gnn_trkorigin}.
The second auxiliary task is vertex prediction, which aims to cluster tracks originating from the same spatial vertex. Ground-truth vertex labels are generated using hierarchical clustering of Monte Carlo truth-level vertex positions, based on a spatial separation threshold of $100~\mu\mathrm{m}$. This threshold reflects the typical position resolution of the tracking detectors and is chosen to ensure practical relevance for $b$-jet identification.
These two auxiliary tasks serve to provide explicit guidance to the model during training, helping it to learn features that are indirectly but critically important for effective $b$-jet tagging. Further details on the training procedure are provided in Section~\ref{sec:GNNs_Training}.

\begin{table}[!htb]
\centering
\begin{tabular}{ll}
\hline
\textbf{Track origin} & \textbf{Description} \\
\hline
Background & Background particles from AMPT  \\
FromC & Charm quark or hadron decay \\
FromB & Beauty quark or hadron decay \\
Primary & Primary vertex except for $c$/$b$ origins \\
OtherSecondary & Secondary vertices except for $c$/$b$ origins \\
\hline
\end{tabular}
\caption{Track origin categories of the GNN model.}
\label{tab:gnn_trkorigin}
\end{table}

\subsection{Training}
\label{sec:GNNs_Training}

% Training procedure
The GNN models are trained separately on pure pp jets and on \PbPb background-overlaid jets for each of the five centrality classes (0--20\%, 20--40\%, 40--60\%, 60--80\%, and 80--100\%). In total, 3.63 million pp jets are used for training and validation, with a $0.8:0.2$ split. Each dataset follows a class composition of approximately $1:2:25$ for \bjets, $c$-jets, and light-flavour jets, respectively. For centrality-dependent training, each pp jet in the dataset is overlaid with background particles from \PbPb events corresponding to the target centrality class.
Table~\ref{tab:gnn_training} summarises the hyperparameters used in training. To reduce memory usage and avoid Out-Of-Memory (OOM) errors, a batch size of 400 is used when training on background-overlaid jets, while a larger batch size of 1000 is adopted for pp jet training.

\begin{table}[!htb]
\centering
\begin{tabular}{ll}
\hline
\textbf{Trainable parameters} & 819~434 \\
\textbf{Optimiser} & Adam \\
\textbf{Learning rate} & 0.001 \\
\textbf{Batch size} & 1000 (pp jets) \\
 & 400 (background-overlaid jets) \\
\textbf{Epochs} & 100 \\
\hline
\end{tabular}
\caption{The training specifications of the GNN model.}
\label{tab:gnn_training}
\end{table}

% Loss

The model is trained using the same composite loss function as in the ATLAS GN1 study, defined as:
\begin{align}
L_\mathrm{total} = L_\mathrm{jet} + \alpha L_\mathrm{vertex} + \beta L_\mathrm{track},
\label{eq:loss}
\end{align}
where:
\begin{itemize}
    \item $L_\mathrm{jet}$ is the categorical cross-entropy loss for jet flavour classification,
    \item $L_\mathrm{vertex}$ is the binary cross-entropy loss averaged over the predicted connections between track pairs for vertex assignment,
    \item $L_\mathrm{track}$ is the categorical cross-entropy loss for predicting the origin of each track.
\end{itemize}
The loss weights are set to $\alpha = 1.5$ and $\beta = 0.5$. To mitigate the effects of class imbalance in both jet flavour and track origin labels, class weights inversely proportional to their respective frequencies are applied during training.
This multi-objective loss ($L_\mathrm{total}$) design allows the network not only to optimise jet flavour classification but also to benefit from auxiliary tasks that enhance its sensitivity to features relevant to flavour tagging.

\begin{figure*}[!htb]
\centering
  \includegraphics[width=0.49\linewidth]{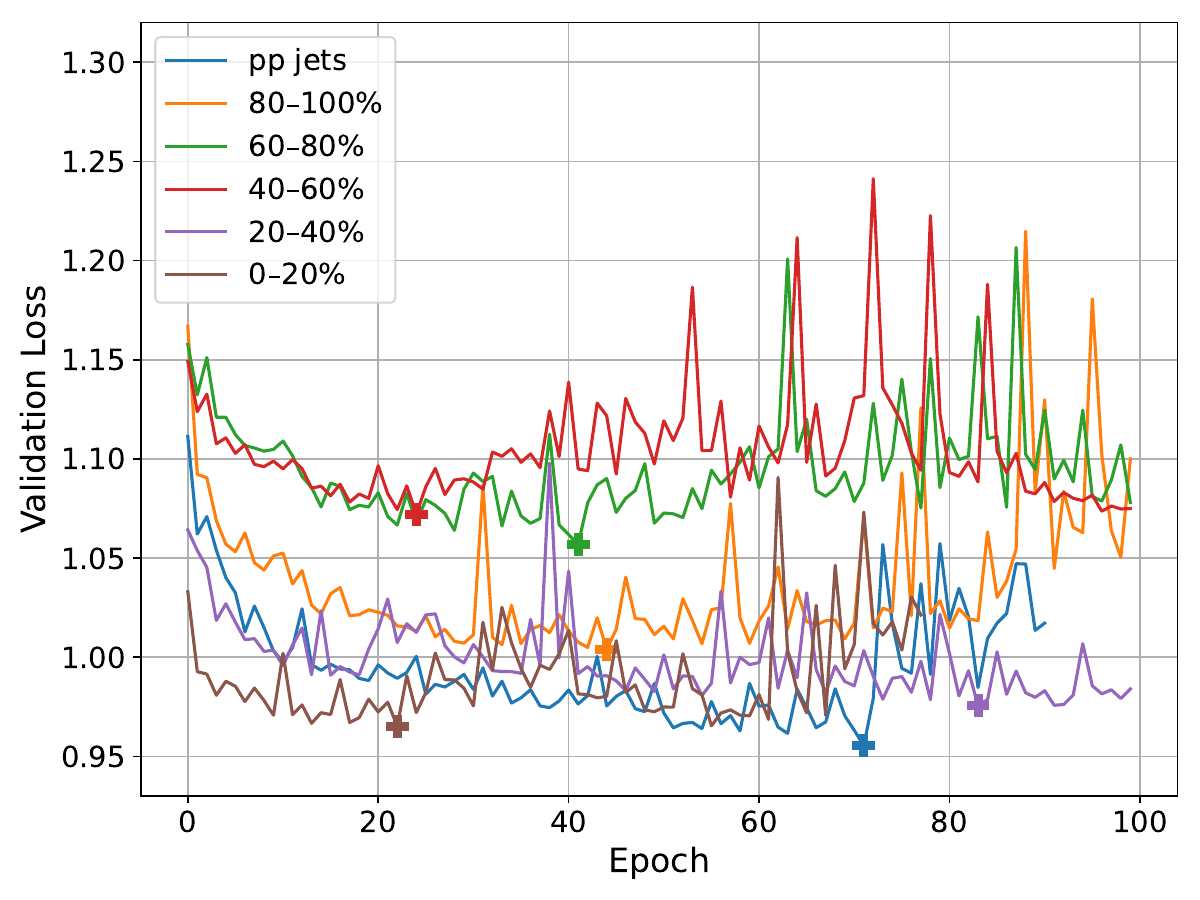}
  \includegraphics[width=0.49\linewidth]{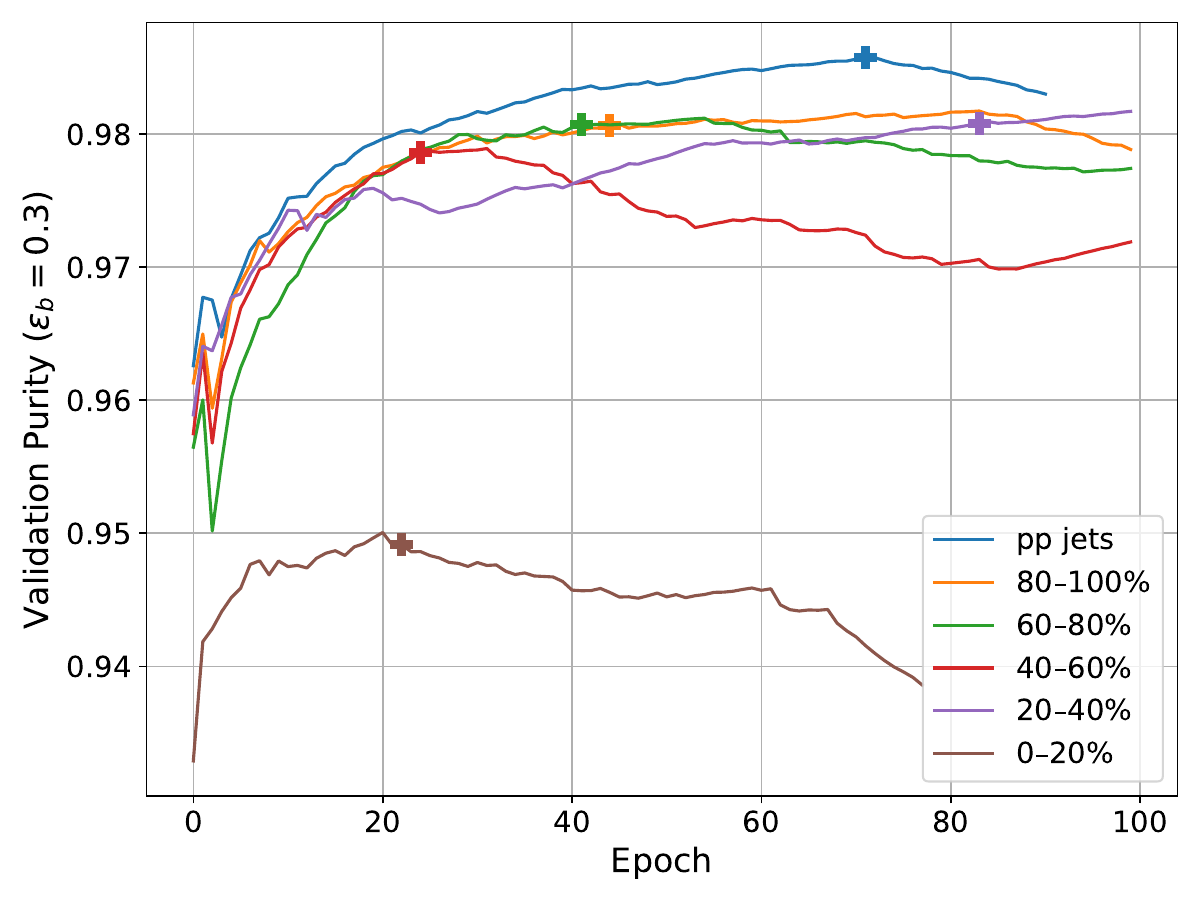}
\caption{Validation loss (left) and $b$-jet tagging purity at the working point of $\epsilon_b = 0.3$ (right) as functions of training epochs. The cross marker indicates the model corresponding to the lowest validation loss, selected as the final model.}
\label{fig:gnn_train}
\end{figure*}

Figure~\ref{fig:gnn_train} shows the evolution of the validation loss and \bjet tagging purity at a fixed working point of $\epsilon_b = 0.3$ throughout training. The model corresponding to the minimum validation loss is selected as the final version to avoid overfitting.
Training times per epoch range from approximately 2 minutes for pure pp jets to 15 minutes for jets with 0--20\% centrality background, using 15 workers and an NVIDIA GeForce RTX 4080 SUPER GPU accelerator.

\section{Results}
\label{sec:Results}

\subsection{Impact of the \PbPb background}
\label{sec:Results_Impact}

% Impact of PbPb backgrounds on GNN performance & tagging decision ($D_b$)
To evaluate the influence of heavy-ion background on tagging performance, the GNN model trained on pp jets is applied to both pure pp jets and their counterparts overlaid with \PbPb background particles. This enables a direct assessment of the model’s response in the presence versus absence of heavy-ion backgrounds.

% b-jet selection
The trained GNN assigns a jet-level probability to each flavour class—$p_b$, $p_c$, and $p_\mathrm{lf}$—corresponding to $b$-, $c$-, and light-flavour jets, respectively. To identify \bjets, a discriminant variable \db is computed from these probabilities, defined as:
\begin{align}
D_{b} = \log \frac{p_b}{(1 - f_c) p_\mathrm{lf} + f_c p_c},
\label{eq:db}
\end{align}
where $f_c$ is a tunable parameter that controls the relative weighting between $c$- and light-flavour jets in the denominator~\cite{ATLASGN1}. In this study, $f_c$ is set to 0.018.

\begin{figure}[!htb]
\centering
  \includegraphics[width=0.49\linewidth]{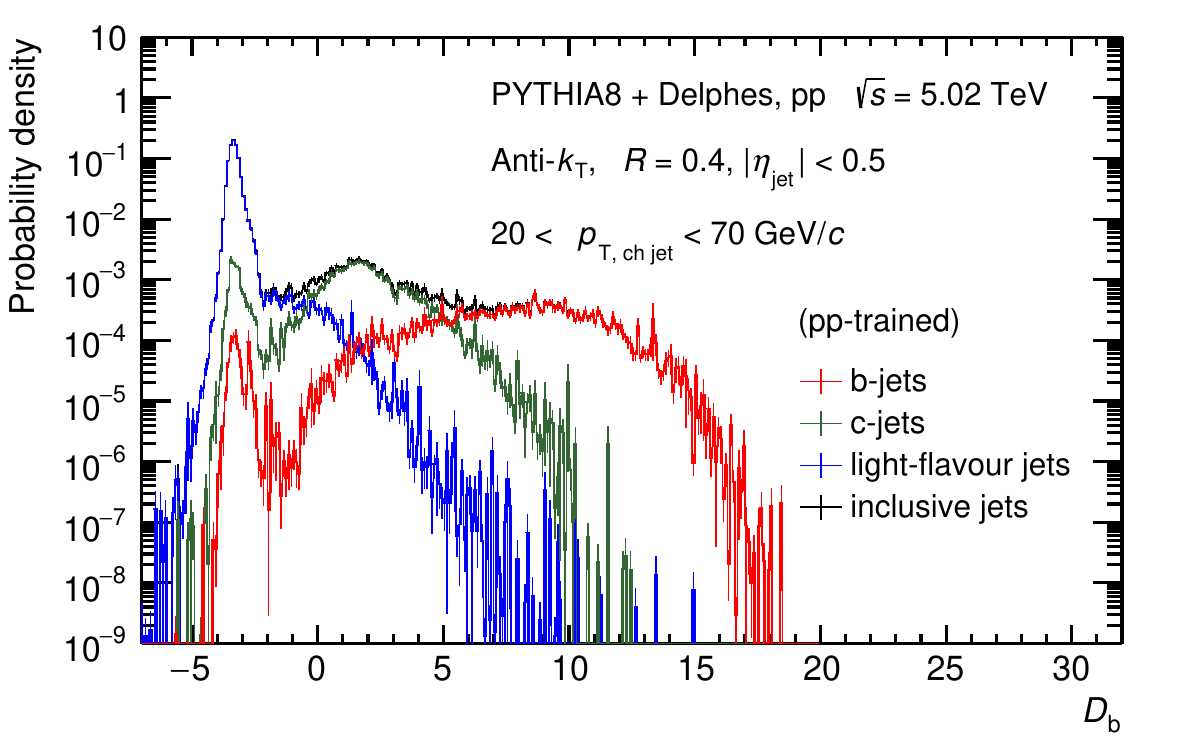}
  \includegraphics[width=0.49\linewidth]{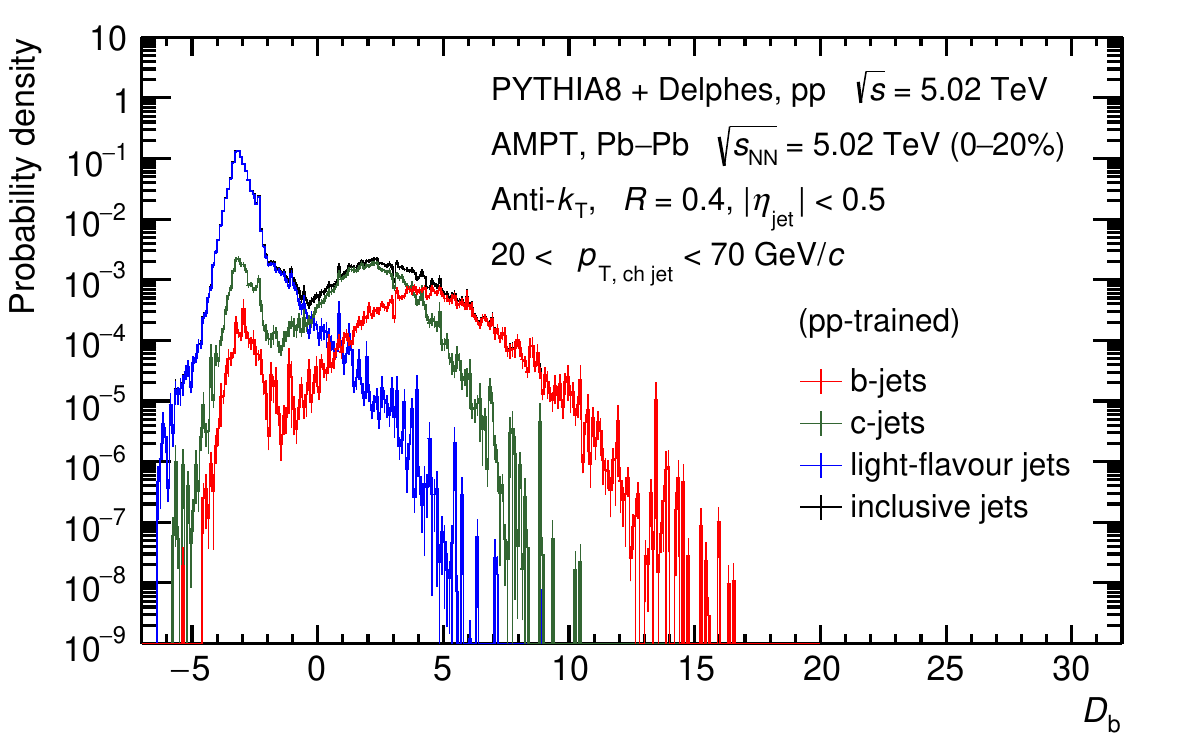}
  \includegraphics[width=0.49\linewidth]{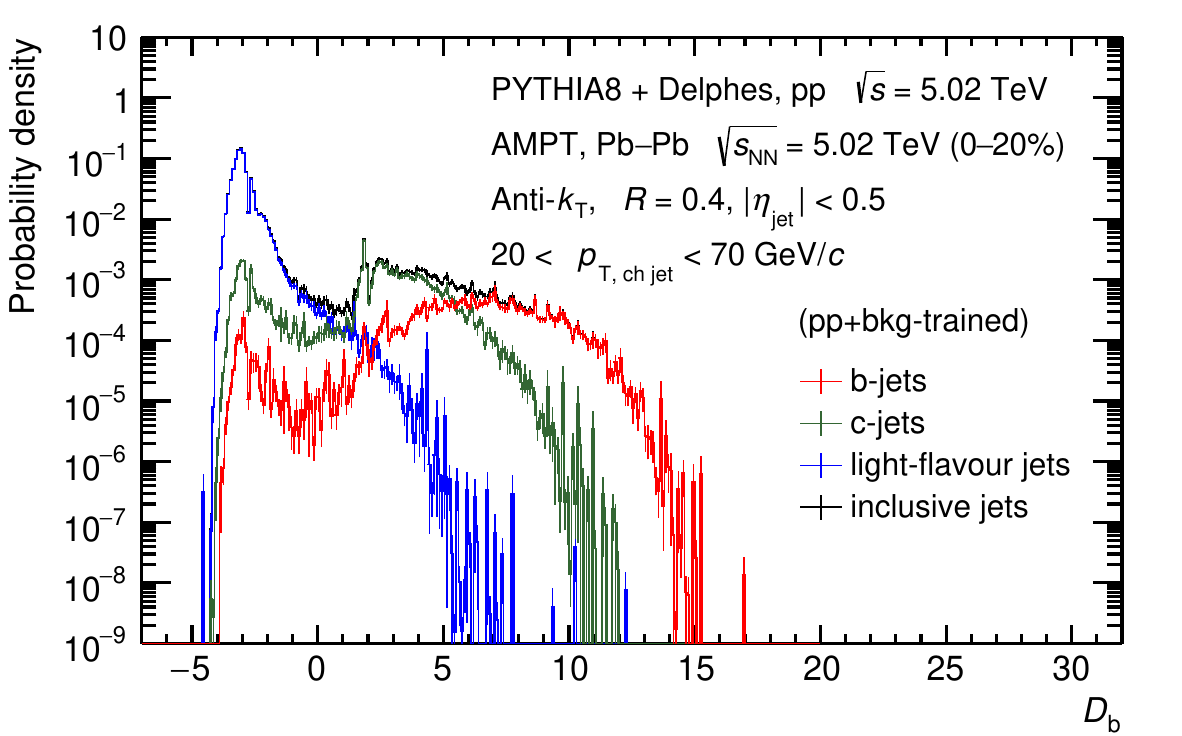}
\caption{Distributions of the \db variable for pp jets and for jets overlaid with 0--20\% \PbPb background. Results are shown for two GNN models: one trained on pp jets only (denoted pp-trained) and the other trained on 0--20\% background-overlaid jets (denoted pp+bkg-trained).}
\label{fig:db_pbpb}
\end{figure}

\begin{figure*}[!htb]
\centering
  \includegraphics[width=0.325\linewidth]{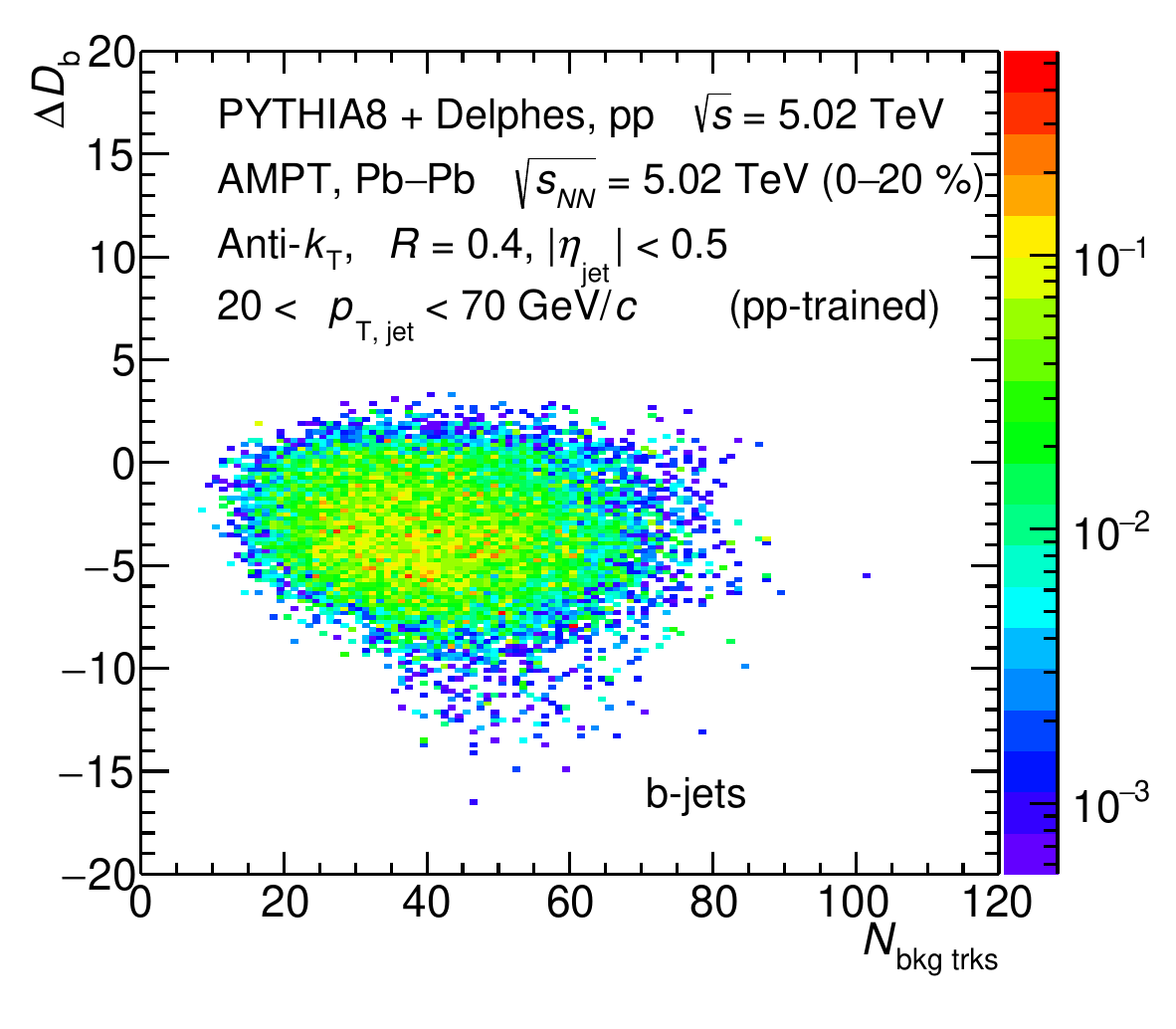}
  \includegraphics[width=0.325\linewidth]{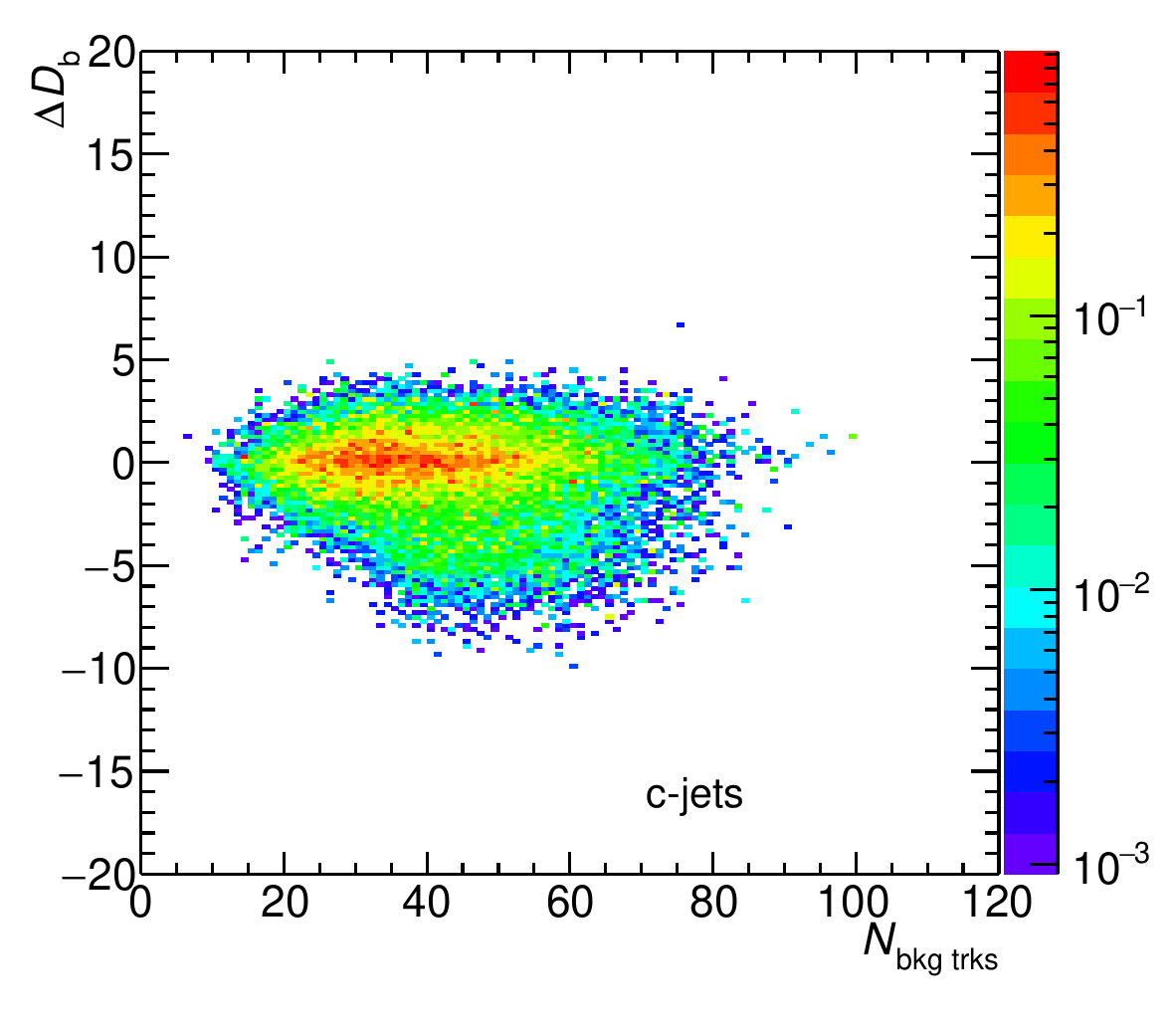}
  \includegraphics[width=0.325\linewidth]{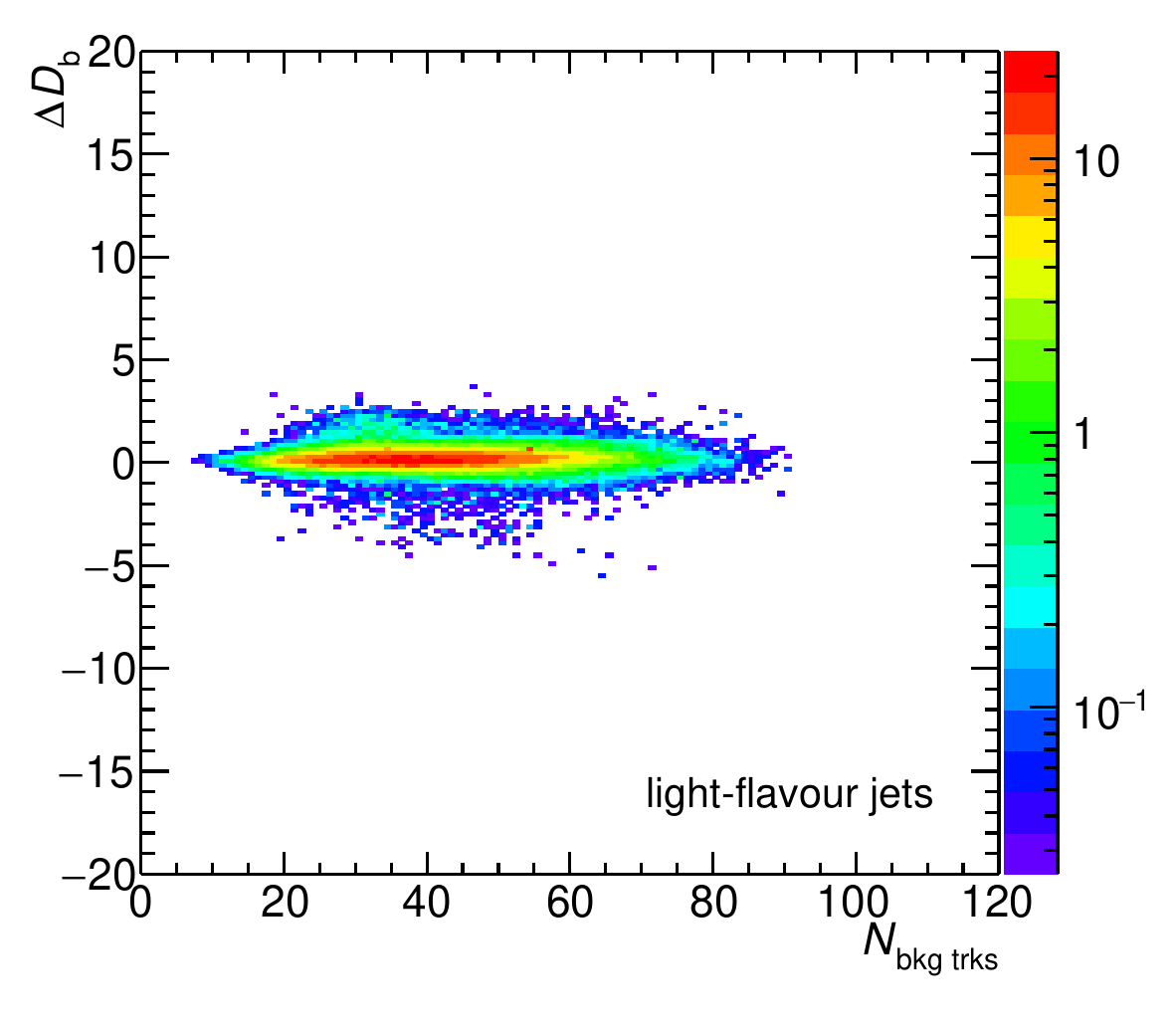}
  \includegraphics[width=0.325\linewidth]{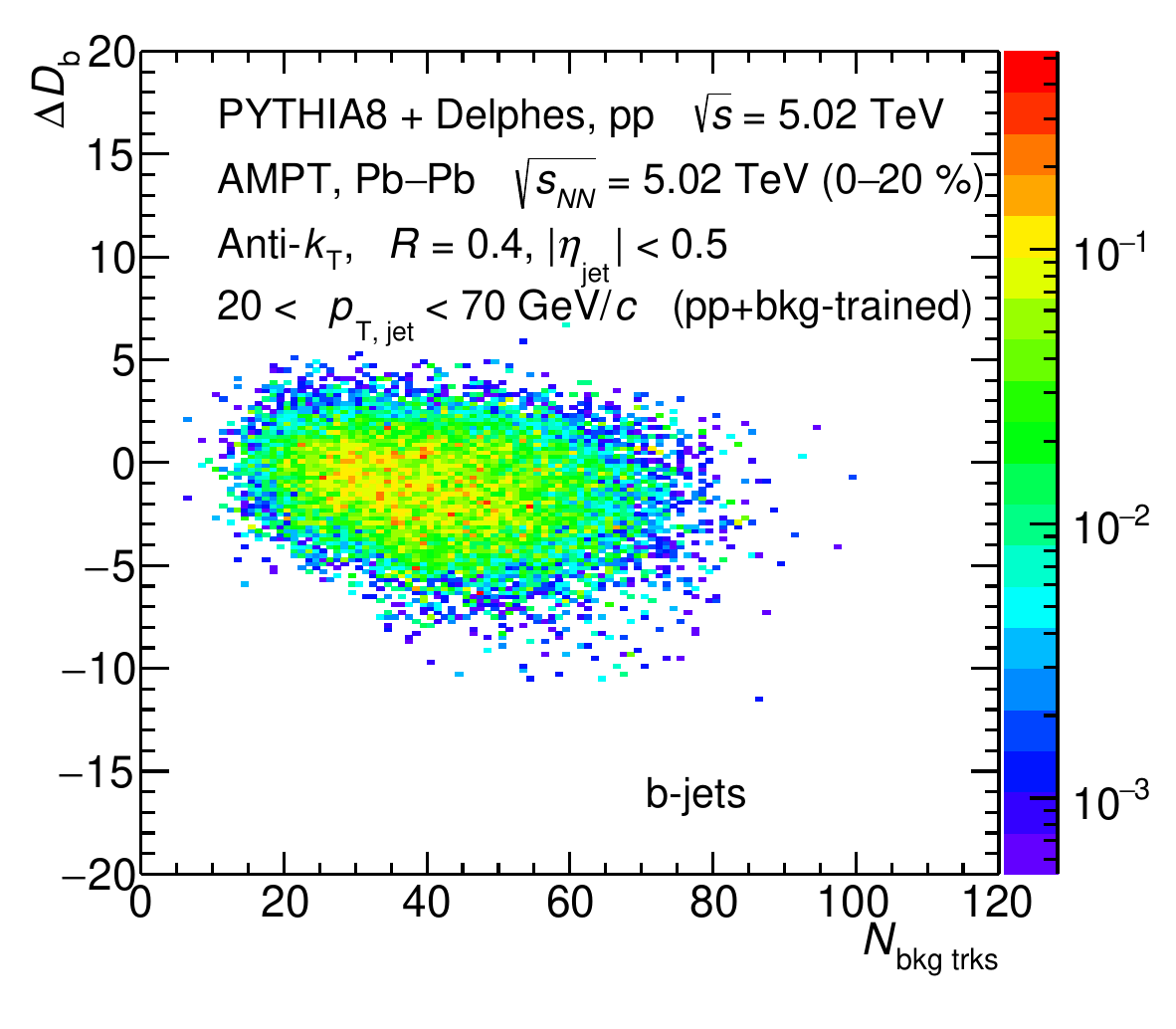}
  \includegraphics[width=0.325\linewidth]{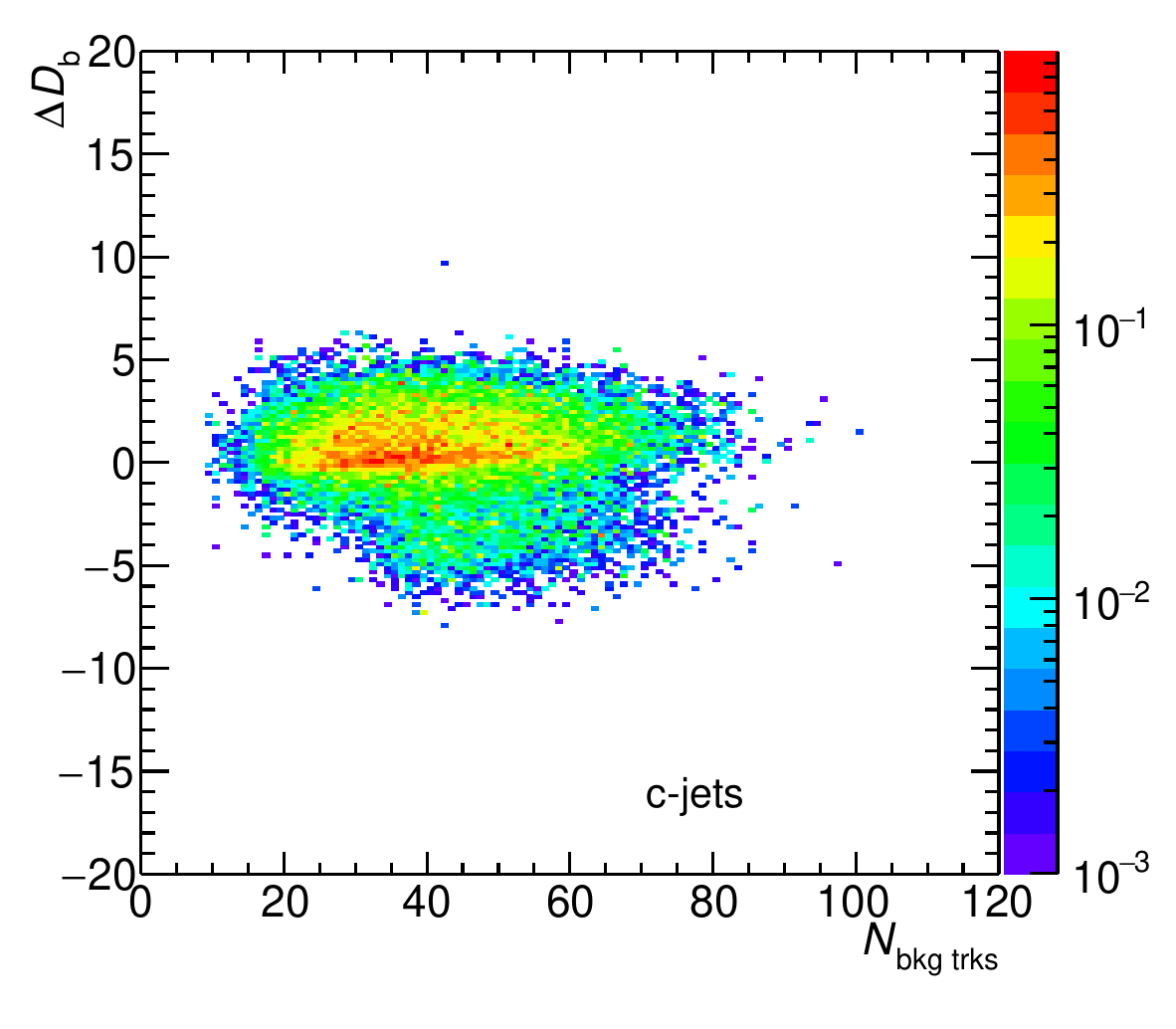}
  \includegraphics[width=0.325\linewidth]{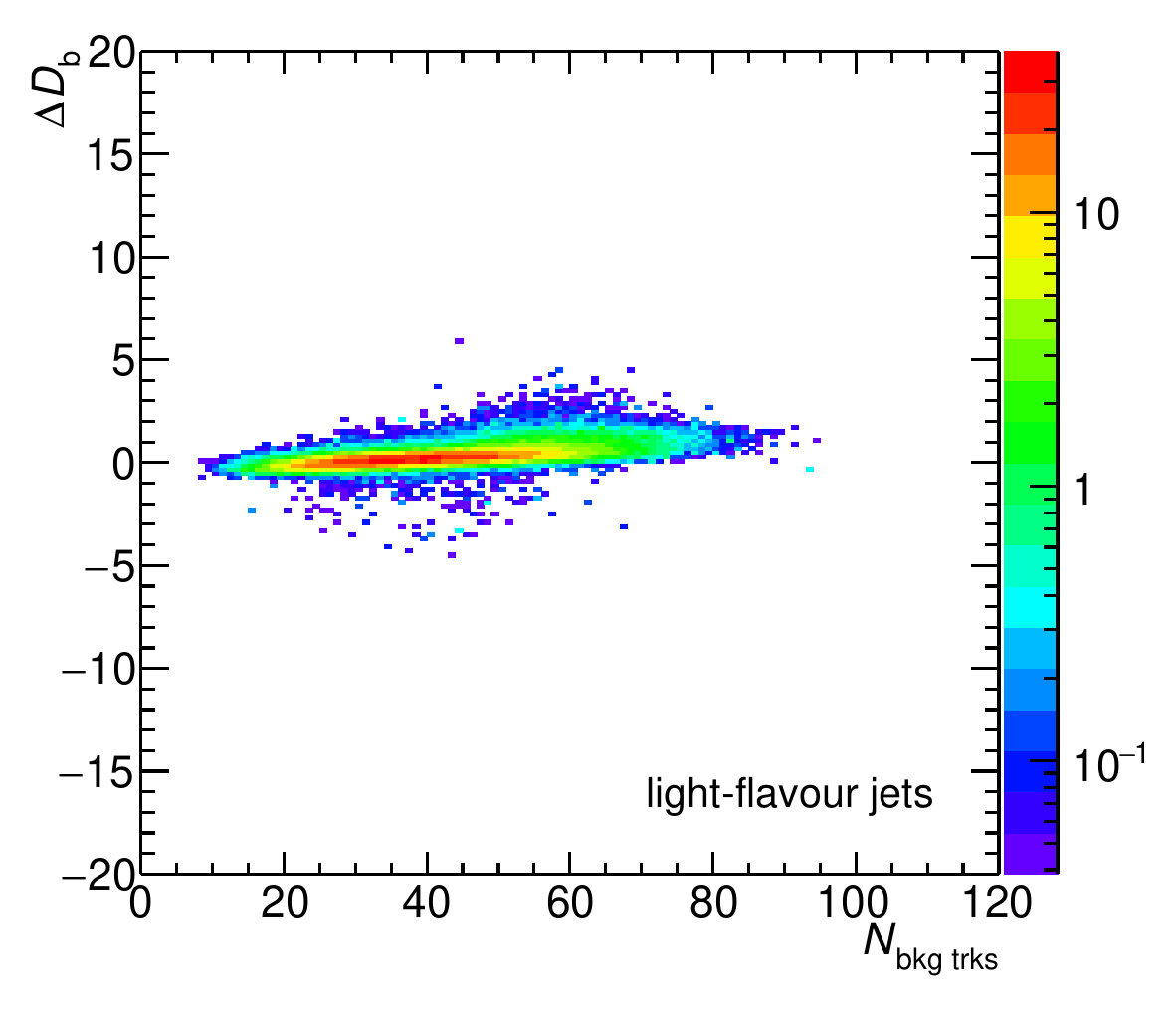}
\caption{Distributions of \db and \ptjet for each jet flavour in the presence of 0--20\% centrality background particles, as predicted by the GNN model trained on pp jets (top) and background-overlaid jets (bottom).}
\label{fig:vardb_pbpb}
\end{figure*}

Figure~\ref{fig:db_pbpb} presents the \db distributions for pure pp jets and 0--20\% background-overlaid jets. For the background-overlaid jets, two different training samples were used—pure pp jets and background-overlaid jets—to assess the impact of the training environment on flavour-tagging performance. In the case of both training and validation on pure pp jets, the \db distribution shows the expected separation: light-flavour jets are concentrated at lower \db values, \bjets at higher values, and $c$-jets in between. This clear separation allows for effective \bjet identification using a threshold on \db.
When background particles from central \PbPb collisions are included, the \db distributions become somewhat narrower, indicating reduced discriminative power. However, the separation among jet flavours remains distinct, suggesting that the model retains its tagging capability even under high-background conditions.

To further examine the effects of heavy-ion background, Fig.~\ref{fig:vardb_pbpb} shows the variations in transverse momentum and \db before and after the overlay of 0--20\% background for each jet. The top and bottom panels correspond to models trained on pure pp jets and background-overlaid jets, respectively. These distributions indicate a systematic decrease in \db for heavy-flavour jets—especially \bjets—when background particles are present. This downward shift in \db implies that the model increasingly misclassifies heavy-flavour jets as light-flavour-like, thereby reducing tagging performance.
Notably, the extent of this \db shift is slightly mitigated when the model is trained on background-overlaid jets, suggesting improved robustness to background contamination. Overall, these findings highlight that heavy-ion background tends to obscure the distinguishing features of heavy-flavour jets, reinforcing the importance of background-aware training for accurate and reliable \bjet tagging in \PbPb collisions.

%\subsection{Background-specific models}
%\label{sec:Results_Specific}

\begin{figure}[!htb]
    \centering
    \includegraphics[width=0.4\linewidth]{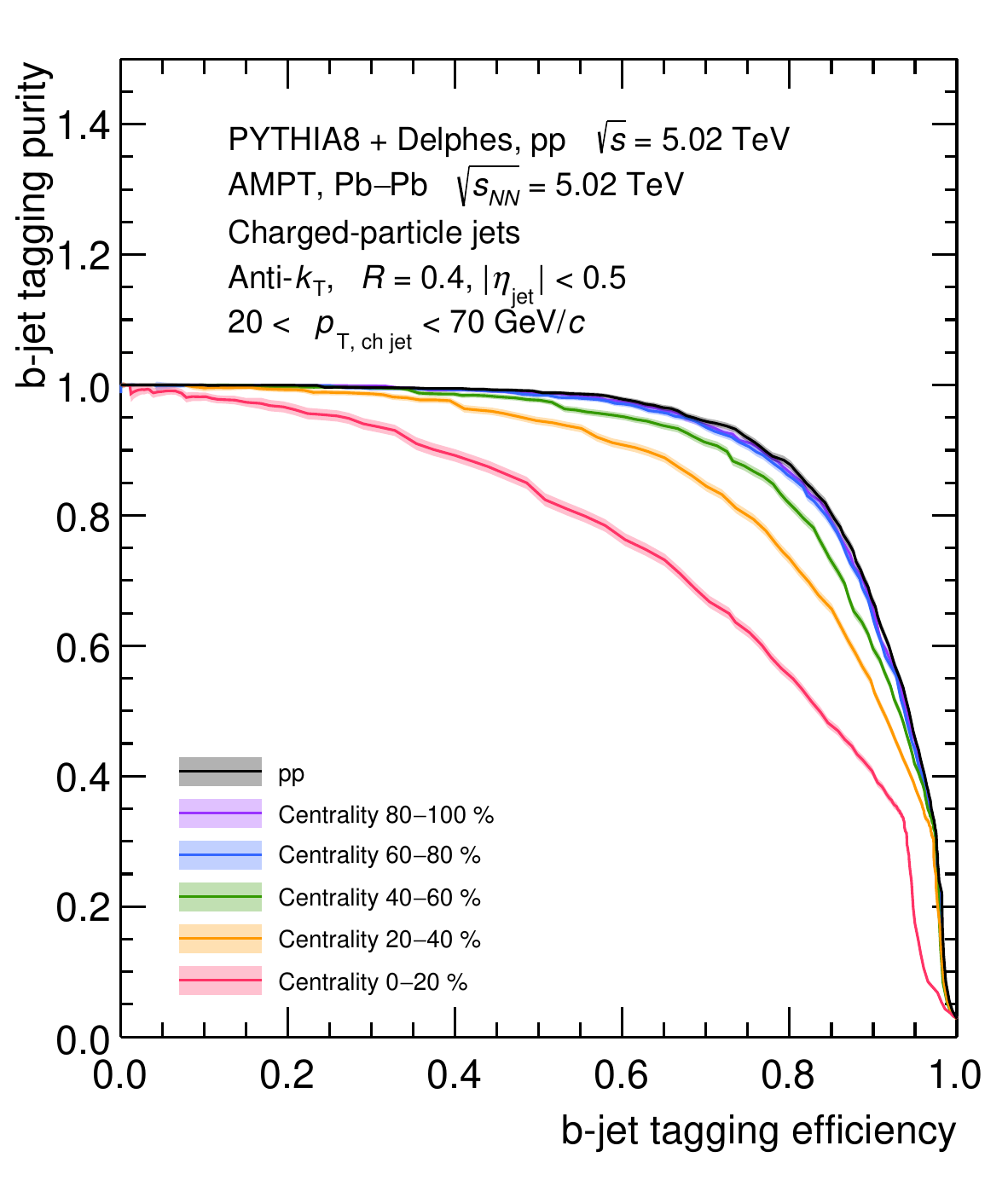}
    \includegraphics[width=0.4\linewidth]{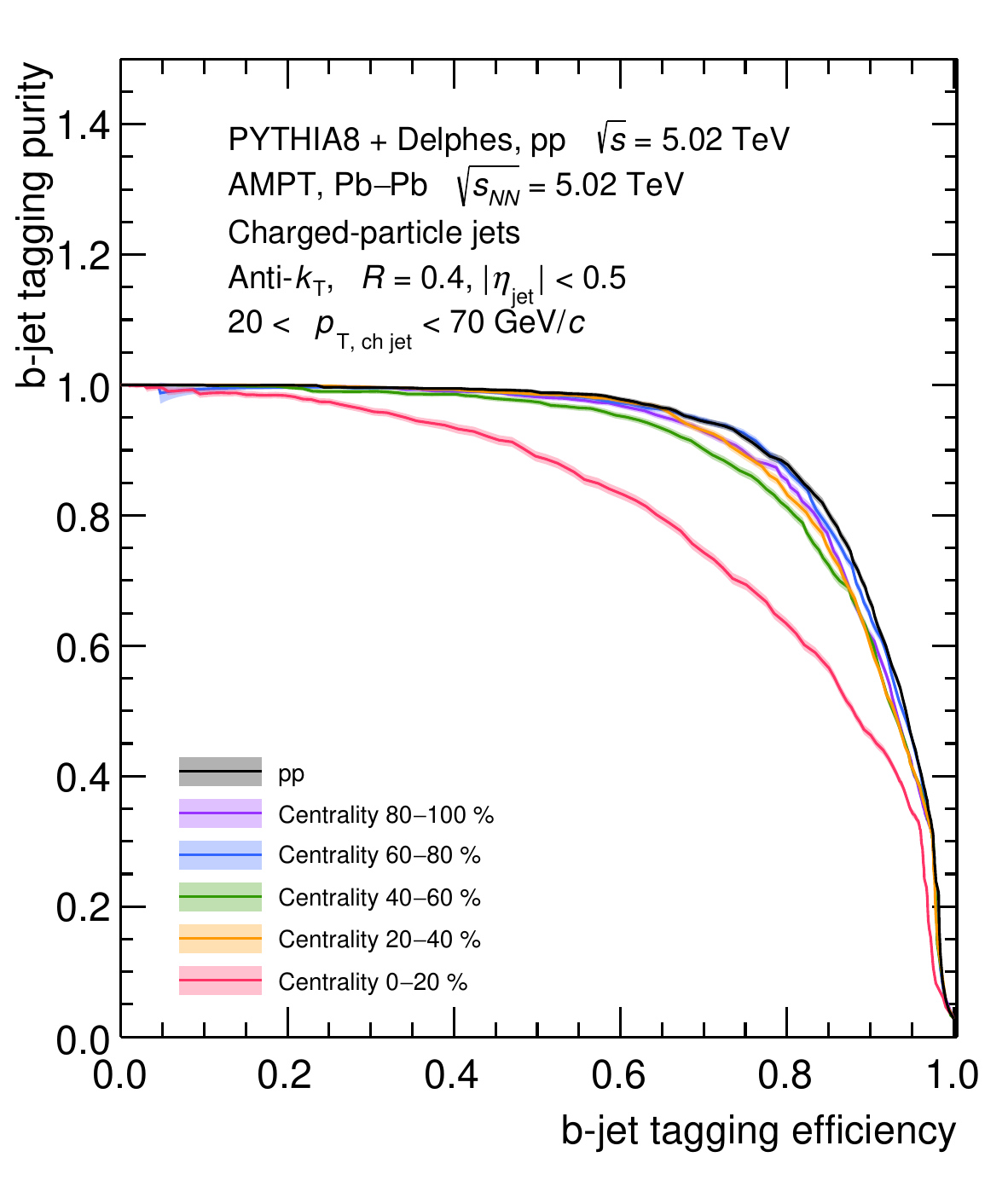}
    \caption{\bjet tagging efficiency and purity obtained using the GNN-based model trained on pure pp jets (left) and on background-overlaid jets (right). The comparison illustrates the impact of training sample composition on tagging performance in the presence of heavy-ion backgrounds.}
    \label{fig:effpur_pbpb}
\end{figure}

Figure~\ref{fig:effpur_pbpb} shows the \bjet tagging performance across different \PbPb centrality classes, evaluated using a validation set of \PbPb background-overlaid jets. The left panel presents results from a model trained on pure pp jets, while the right panel corresponds to a model trained on background-overlaid jets.

As expected, the model trained on pp jets exhibits a progressive decline in tagging performance with increasing centrality, due to the higher background particle density in more central collisions. Nevertheless, the GNN-based \bjet tagging method maintains strong performance, remaining competitive with—or even exceeding—the classical approaches used in pp environments. For instance, the Secondary Vertex (SV) method achieved a tagging efficiency of approximately 0.2–0.3 and a purity of 0.3–0.5 for $20 < p_\mathrm{T,~jet}<70~\mathrm{GeV}/c$ in pp collisions at $\sqrt{s_\mathrm{NN}} = 5.02$ TeV~\cite{ALICEbjet}. In comparison, the GNN model yields significantly better performance, even under the most central \PbPb background conditions.

The performance degradation observed for background-overlaid jets is primarily due to the model’s lack of exposure to background effects during training. To mitigate this, dedicated GNN models were trained for each centrality class using corresponding background-overlaid jets. The results, shown in the bottom panels of Figure~\ref{fig:effpur_pbpb}, demonstrate a clear improvement in tagging performance when the training incorporates centrality-specific background. These findings underscore the importance of background-aware training in achieving reliable \bjet identification in heavy-ion collisions.

\begin{figure}[!htb]
    \centering
    \includegraphics[width=0.55\linewidth]{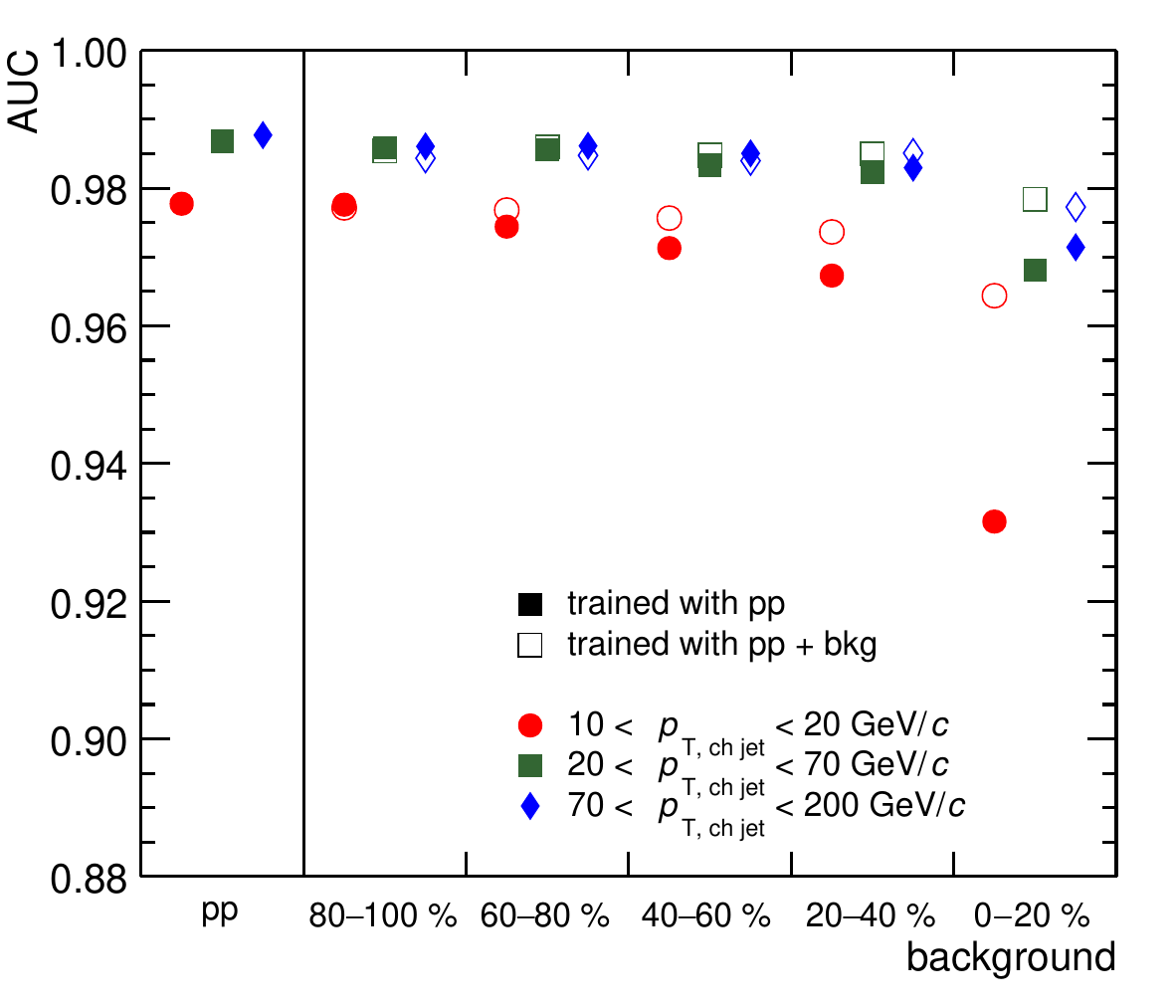}
    \caption{Receiver Operating Characteristic (ROC) Area Under Curve (AUC) values of the GNN-based $b$-jet tagging model trained on pp jets (shown with filled markers) and on background-overlaid jets (shown with open markers), evaluated for each centrality class of the \PbPb background events.}
    \label{fig:auc_pbpb}
\end{figure}

Figure~\ref{fig:auc_pbpb} presents the performance of GNN-based $b$-jet tagging models trained on either pure pp jets or background-overlaid jets. To provide a comprehensive evaluation of each model’s classification capability across all possible threshold values, we adopt the Area Under the Curve (AUC) of the Receiver Operating Characteristic (ROC) curve—a widely used metric in machine learning. The ROC curve displays the true positive rate (TPR) as a function of the false positive rate (FPR), with a diagonal line ($x = y$) representing the performance of a random classifier. The ideal point, denoted by FPR = 0 and TPR = 1, represents perfect classification. The ROC AUC values range from 0 to 1, with higher values indicating stronger classification performance in terms of both efficiency and purity.

As shown in the figure, the model trained exclusively on pp jets exhibits a clear decline in performance with increasing \PbPb centrality, reflecting the growing complexity introduced by higher background particle multiplicities. This degradation is especially pronounced for low-\ptrm jets, where background contamination has a greater impact.
By contrast, models trained on background-overlaid jets—each tailored to a specific centrality class—show a similar centrality-dependent trend, but with a significantly reduced degradation. In particular, for the most central background events, the ROC AUC values remain noticeably higher than those of the pp-trained model, demonstrating that incorporating background conditions during training improves the model’s robustness and maintains tagging performance even in challenging heavy-ion environments.

\section{Summary}
\label{sec:Sum}

To enable precise measurements of low-\ptrm \bjets in heavy-ion collisions, we performed a study of utilizing a \bjet tagging method optimised for the ALICE detector environment. In this study, we adapted GN1--a graph neural network (GNN)--based tagging algorithm originally developed by the ATLAS Collaboration, for use in ALICE analyses. As a preliminary step, we investigated the influence of \PbPb background particles on the GNN’s tagging performance by evaluating the GN1 model, trained on \pythia-generated pp jets, using jets overlaid with background particles from AMPT \PbPb simulations.
This approach does not incorporate medium-induced effects such as jet energy loss, deformation, or the presence of long-lived background particles that may produce secondary vertices. Consequently, it has inherent limitations in reproducing the full complexity of heavy-ion environments. However, it offers a controlled framework for qualitatively assessing the impact of the dense particle background characteristic of \PbPb collisions on GNN-based \bjet tagging.
The study reveals that the inclusion of central \PbPb background, corresponding to high particle multiplicity, degrades the tagging performance. In particular, as the number of associated background particles increases, the GNN tends to shift its classification toward light-flavour jets. Despite this degradation, the GNN model continues to outperform traditional tagging methods. Moreover, training dedicated models with background samples specific to each centrality class significantly mitigates the performance loss.
These findings suggest that GNN-based tagging approaches hold strong potential for future \bjet analyses in ALICE \PbPb collisions, especially when trained with realistic heavy-ion background conditions.

\section*{Acknowledgments}
%\acknowledgments
%\textit{Acknowledgments.---} 
C. Choi and S. Lim are supported by the National Research Foundation of Korea (NRF) grant funded by the Korea government (MSIT) under Contract No. RS-2008-NR007226 and RS-2025-00554431. 
We also acknowledge technical support from KIAF administrators at KISTI.

\section*{References}
\bibliographystyle{iopart-num} 
\bibliography{main}

\providecommand{\newblock}{}
\begin{thebibliography}{10}
\expandafter\ifx\csname url\endcsname\relax
  \def\url#1{{\tt #1}}\fi
\expandafter\ifx\csname urlprefix\endcsname\relax\def\urlprefix{URL }\fi
\providecommand{\eprint}[2][]{\url{#2}}

\bibitem{Busza:2018rrf}
Busza W, Rajagopal K and van~der Schee W 2018 {\em Ann. Rev. Nucl. Part. Sci.\/} {\bf 68} 339--376 (\textit{Preprint} \eprint{1802.04801})

\bibitem{PDG}
Navas S {\em et~al.\/} (Particle Data Group) 2024 {\em Phys. Rev. D\/} {\bf 110} 030001

\bibitem{Dong:2019byy}
Dong X, Lee Y~J and Rapp R 2019 {\em Ann. Rev. Nucl. Part. Sci.\/} {\bf 69} 417--445 (\textit{Preprint} \eprint{1903.07709})

\bibitem{Dokshitzer:2001zm}
Dokshitzer Y~L and Kharzeev D~E 2001 {\em Phys. Lett. B\/} {\bf 519} 199--206 (\textit{Preprint} \eprint{hep-ph/0106202})

\bibitem{ALICEdeadcone}
{ALICE Collaboration} (ALICE) 2022 {\em Nature\/} {\bf 605} 440--446 [Erratum: Nature 607, E22 (2022)] (\textit{Preprint} \eprint{2106.05713})

\bibitem{Zigic:2018ovr}
Zigic D, Salom I, Auvinen J, Djordjevic M and Djordjevic M 2019 {\em Phys. Lett. B\/} {\bf 791} 236--241 (\textit{Preprint} \eprint{1805.04786})

\bibitem{Wicks:2005gt}
Wicks S, Horowitz W, Djordjevic M and Gyulassy M 2007 {\em Nucl. Phys. A\/} {\bf 784} 426--442 (\textit{Preprint} \eprint{nucl-th/0512076})

\bibitem{Ke:2018tsh}
Ke W, Xu Y and Bass S~A 2018 {\em Phys. Rev. C\/} {\bf 98} 064901 (\textit{Preprint} \eprint{1806.08848})

\bibitem{ATLASbjetRAA}
Aad G {\em et~al.\/} (ATLAS) 2023 {\em Eur. Phys. J. C\/} {\bf 83} 438 (\textit{Preprint} \eprint{2204.13530})

\bibitem{CMSbjetRAA}
{CMS Collaboration} (CMS Collaboration) 2014 {\em Phys. Rev. Lett.\/} {\bf 113}(13) 132301 \urlprefix\url{https://link.aps.org/doi/10.1103/PhysRevLett.113.132301}

\bibitem{ALICEreview}
{ALICE Collaboration} 2024 {\em European Physical Journal C\/} {\bf 84} ISSN 1434-6044 publisher Copyright: {\textcopyright} The Author(s) 2024.

\bibitem{ALICEupgrade}
{ALICE Collaboration} 2024 {\em Journal of Instrumentation\/} {\bf 19} P05062 \urlprefix\url{https://dx.doi.org/10.1088/1748-0221/19/05/P05062}

\bibitem{CMS:2012feb}
Chatrchyan S {\em et~al.\/} (CMS) 2013 {\em JINST\/} {\bf 8} P04013 (\textit{Preprint} \eprint{1211.4462})

\bibitem{CMSJPPbPb}
Nguyen M 2013 {\em Nuclear Physics A\/} {\bf 904-905} 705c--708c ISSN 0375-9474 the Quark Matter 2012 \urlprefix\url{https://www.sciencedirect.com/science/article/pii/S0375947413002388}

\bibitem{ATLAS:2015thz}
Aad G {\em et~al.\/} (ATLAS) 2016 {\em JINST\/} {\bf 11} P04008 (\textit{Preprint} \eprint{1512.01094})

\bibitem{ALICEbjet}
{ALICE Collaboration} (ALICE) 2022 {\em JHEP\/} {\bf 2201} 178 35 pages, 16 captioned figures, 2 tables, authors from page 30, published version, figures at http://alice-publications.web.cern.ch/node/7411 (\textit{Preprint} \eprint{2110.06104}) \urlprefix\url{https://cds.cern.ch/record/2783306}

\bibitem{ATLASGN1}
{ATLAS Collaboration} (ATLAS) 2022 {Neural Network Jet Flavour Tagging with the Upgraded ATLAS Inner Tracker Detector at the High-Luminosity LHC} Tech. rep. CERN Geneva all figures including auxiliary figures are available at https://atlas.web.cern.ch/Atlas/GROUPS/PHYSICS/PUBNOTES/ATL-PHYS-PUB-2022-047 \urlprefix\url{https://cds.cern.ch/record/2839913}

\bibitem{GATv2}
Brody S, Alon U and Yahav E 2021  (\textit{Preprint} \eprint{2105.14491})

\bibitem{Mondal:2024nsa}
Mondal S and Mastrolorenzo L 2024 {\em Eur. Phys. J. ST\/} {\bf 233} 2657--2686 (\textit{Preprint} \eprint{2404.01071})

\bibitem{ATLAS:2022qxm}
Aad G {\em et~al.\/} (ATLAS) 2023 {\em Eur. Phys. J. C\/} {\bf 83} 681 (\textit{Preprint} \eprint{2211.16345})

\bibitem{PYTHIA8}
Bierlich C, Chakraborty S, Desai N, Gellersen L, Helenius I, Ilten P, Lönnblad L, Mrenna S, Prestel S, Preuss C~T, Sjöstrand T, Skands P, Utheim M and Verheyen R 2022 A comprehensive guide to the physics and usage of pythia 8.3 (\textit{Preprint} \eprint{2203.11601}) \urlprefix\url{https://arxiv.org/abs/2203.11601}

\bibitem{Delphes}
de~Favereau J, Delaere C, Demin P, Giammanco A, Lema\^\i{}tre V, Mertens A and Selvaggi M (DELPHES 3) 2014 {\em JHEP\/} {\bf 02} 057 (\textit{Preprint} \eprint{1307.6346})

\bibitem{GEANT4:2002zbu}
Agostinelli S {\em et~al.\/} (GEANT4) 2003 {\em Nucl. Instrum. Meth. A\/} {\bf 506} 250--303

\bibitem{FastJet}
Cacciari M, Salam G~P and Soyez G 2012 {\em Eur. Phys. J. C\/} {\bf 72} 1896 (\textit{Preprint} \eprint{1111.6097})

\bibitem{AntikT}
Cacciari M, Salam G~P and Soyez G 2008 {\em Journal of High Energy Physics\/} {\bf 2008} 063 \urlprefix\url{https://dx.doi.org/10.1088/1126-6708/2008/04/063}

\bibitem{AMPT}
Lin Z~W, Ko C~M, Li B~A, Zhang B and Pal S 2005 {\em Phys. Rev. C\/} {\bf 72} 064901 (\textit{Preprint} \eprint{nucl-th/0411110})

\bibitem{ALICE:2013hur}
Abelev B {\em et~al.\/} (ALICE) 2013 {\em Phys. Rev. C\/} {\bf 88} 044909 (\textit{Preprint} \eprint{1301.4361})

\bibitem{PyTorch}
Paszke A, Gross S, Massa F, Lerer A, Bradbury J, Chanan G, Killeen T, Lin Z, Gimelshein N, Antiga L, Desmaison A, Köpf A, Yang E, DeVito Z, Raison M, Tejani A, Chilamkurthy S, Steiner B, Fang L, Bai J and Chintala S 2019 Pytorch: An imperative style, high-performance deep learning library (\textit{Preprint} \eprint{1912.01703}) \urlprefix\url{https://arxiv.org/abs/1912.01703}

\bibitem{PyTorchL}
Falcon W and {The PyTorch Lightning team} 2019 {PyTorch Lightning} \urlprefix\url{https://github.com/Lightning-AI/lightning}

\end{thebibliography}

\end{document}